\documentclass[twocolumn,showpacs,preprintnumbers,amsmath,amssymb,superscriptaddress,prb]{revtex4}
\usepackage{graphicx}% Include figure files
\usepackage{dcolumn}% Align table columns on decimal point
\usepackage{bm}% bold math
\usepackage{subfigure}
\usepackage{color}
\usepackage{amsfonts}%
\usepackage{amssymb}
\usepackage{amsmath}%
\begin{document}

%\preprint{APS/123-QED}

\title{Terahertz Hall Measurements On Optimally Doped Single Crystal $\rm Bi_2 Sr_2 Ca Cu_2 O_{8+x}$}

\author{G. S. Jenkins}
\author{D. C. Schmadel}%
\author{A. B. Sushkov}
\affiliation{%
Center for Nanophysics and Advanced Materials\\Department of
Physics, University of Maryland, College Park, Maryland 20742, USA
}%

\author{G. D. Gu}
\affiliation{Condensed Matter Physics \& Materials Science
Department, Brookhaven National Laboratory, Upton, NY 11973-5000,
USA}
\author{H. Kontani}
\affiliation{
Department of Physics, Nagoya University, Furo-cho, Nagoya 464-8602, Japan% with \\
}%

\author{H. D. Drew}
\affiliation{%
Center for Nanophysics and Advanced Materials\\Department of
Physics, University of Maryland, College Park, Maryland 20742, USA
}%
\date{\today}% It is always \today, today,

\begin{abstract}

The infrared Hall angle in optimally doped single crystal $\rm
Bi_2 Sr_2 Ca Cu_2 O_{8+x}$  was measured from 3.05 to 21.75 meV as
a continuous function of temperature from 25 to 300\,K.  In the
normal state, the temperature dependence of the real part of the
cotangent of the infrared Hall angle obeys the same power law as
dc measurements. The measured Hall frequency $\rm \omega_H$ is
significantly larger than the expected value based upon ARPES data
analyzed in terms of the relaxation time approximation. This
discrepancy as well as the temperature dependence of $\rm
Re(\cot{\theta_H})$ and $\omega_H$ is well described by a Fermi
liquid theory in which current vertex corrections produced by
electron-magnon scattering are included.

%Valid PACS numbers may be entered using the \verb+\pacs{#1}+ command.
\end{abstract}

\maketitle

\section{Introduction}

Ever since the discovery of High T$_c$ cuprate superconductors,
Hall angle measurements have figured prominently in the discussion
of two broad classes of theoretical approaches of the
nonsuperconducting state: Fermi liquid and non-Fermi liquid
descriptions. The cotangent of the dc-Hall angle in optimally
doped p-type cuprates exhibits an anomalous, nearly quadratic
temperature dependence
\cite{Konstantinovic_dcHall_Bi2212,Ri_dcHall_OptBi2212,Forro_dcHall_OptBi2212,Xu_dcHall_overBi2212,Ong_dcHall_Tsquare}
while the longitudinal resistivity shows a linear dependence over
a wide range of temperature.\cite{Ong_dcHall_Tsquare} The measured
dc-Hall coefficient $R_H$ near the superconducting transition
temperature T$_c$ is a factor of 3 to 4 larger
\cite{Konstantinovic_dcHall_Bi2212, Ri_dcHall_OptBi2212,
Forro_dcHall_OptBi2212, Xu_dcHall_overBi2212} than that predicted
by Luttinger's thereom  while angular resolved photoemission
spectroscopy (ARPES) measurements show a reasonably simple large
holelike Fermi surface (FS) whose area is consistent with the
stoichiometric doping (although there are some subtleties
associated with the bilayer splitting in double-layer
compounds).\cite{Damascelli_RMP_ARPESReview} These facts have
often been interpreted as proof of distinctly non-Fermi liquid
behavior spurning the development of novel and exotic theories to
describe the phenomenology.\cite{AndersonBook,VarmaMFL}

However, recent Shubnikov de Haas (SdH) and de Haas van Alphen
(dHvA) measurements have demonstrated sharp oscillations in 1/H in
both overdoped and underdoped p-type
cuprates,\cite{QO_UD1,QO_UD2,QO_UD3,QO_UD4,Vignolle_QO_Tl2201} a
hallmark signature of quantum oscillations and the existence of a
well defined Fermi surface and attendant quasiparticles. In
overdoped Tl$_2$Ba$_2$CuO$_{6+\delta}$
(Tl-2201),\cite{Vignolle_QO_Tl2201} the area of the FS as deduced
from dHvA oscillations is consistent with ARPES
measurements,\cite{Damascelli_Tl2201_ARPES} angular
magnetoresistance oscillations (AMRO)
measurements,\cite{Hussey-Nature2006} low temperature Hall
coefficient measurements,\cite{mackenzie_Tl_RH} and Luttinger's
thereom.\cite{Damascelli_Tl2201_ARPES} There is general agreement
between ARPES and band structure
calculations.\cite{Peets_AMRO,Singh_DFTBandCalcs_Tl2201}
Importantly, the Wiedemann-Franz law has been experimentally
verified.\cite{Proust_WeidermanFranz_Tl2201Overdoped,Peets_AMRO}
This evidence appears to validate Fermi liquid theory at least in
the strongly overdoped p-type cuprates.

Underdoped p-type cuprates exhibit more complicated behavior.
Measured in zero field at temperatures above T$_c$, the FS
depicted by ARPES shows disconnected gapless Fermi arcs that have
holelike curvature \cite{Damascelli_RMP_ARPESReview} whose length
diminishes with decreasing doping and
temperature.\cite{Kanigel_neutrons_UD} In high field and low
temperature, dc-$R_H$ measurements show a negative electronlike
behavior \cite{taillefer-2009} while the measured frequency of
quantum oscillations in 1/H imply a drastically smaller FS than
the large holelike FS implied by ARPES. The measured value of the
cyclotron mass in underdoped systems, $m_c \sim 3
m_e$,\cite{QO_UD1,QO_UD2,QO_UD3,QO_UD4} is very near the band
value associated with the full FS. Generally for a small FS, one
would expect a large variation in the energy dispersion resulting
in a much reduced cyclotron mass.

Unfortunately, no quantum oscillation experiments have been
performed on optimally doped material since relatively short
lifetimes and the present limit of achievable magnetic fields
conspire to make measurements untenable.  Infrared (IR) Hall
measurements do not suffer from such strict constraints and have
been performed on YBa$_2$Cu$_3$O$_{6+x}$(YBCO-123) thin films at
$\sim$100 meV in low fields ($\sim$8T) above T$_c$, circumstances
more like the zero field ARPES conditions than the high field
quantum oscillations experiments. The IR Hall frequency is
holelike and increases with decreasing doping, a signature of FS
reconstruction consistent with formation of small pockets.
\cite{rigal}  However, the relatively high frequency of the
reported IR Hall measurements complicate direct comparisons with
ARPES measurements due to known intermediate energy scales such as
the high energy renormalization of the dispersion as measured by
ARPES, \cite{Damascelli_RMP_ARPESReview} interband
transitions,\cite{Padilla_Opticalsxx_Constantmxx} the pseudogap
energy,\cite{Timusk_Statt_Psuedogap} and magnetic correlations
measured by neutron experiments.\cite{Fong_neutron_YBCO}
Furthermore, the Cu-O chain contribution to the longitudinal
conductivity uniquely associated with YBCO-123 complicates the
analysis of the IR Hall data making direct quantitative
comparisons difficult.

IR Hall measurements at low frequencies ($\lesssim 10$ meV) below
these characteristic energy scales directly probe the intrinsic
properties of the FS in a unique way. This technique, together
with quantitative comparisons with ARPES and other transport data,
have proven very successful in establishing the causal mechanism
of the extremely anomalous behavior of the Hall effect in n-type
cuprates in the overdoped regime \cite{Jenkins-overdopedPCCO,
Kontani_2008Review} as well as the underdoped regime in the
paramagnetic state.\cite{Jenkins2009PRB} Furthermore, the
technique has proven a reliable and sensitive probe of FS
reconstruction as established in underdoped n-type cuprates below
the N\`eel temperature, \cite{ARPES_edoped_Park,
ARPES_edoped_Armitage} manifesting as a drastically reduced Hall
mass.\cite{Jenkins2009PRB}

In this paper, we report finite frequency Hall angle measurements
on optimally doped $\rm Bi_2 Sr_2 Ca Cu_2 O_{8+x}$  (Bi-2212)
single crystals in the terahertz (THz), or far infrared (FIR),
spectral region. The FIR Hall response is significantly larger
than expected based upon ARPES results analyzed within the
relaxation time approximation (RTA), a result similar to dc Hall
measurements. This discrepancy can be accounted for by including
current vertex corrections produced by electron-magnon scattering
in a Fermi-liquid based theoretical analysis, a mechanism firmly
established in n-type cuprates.
\cite{Kontani_2008Review,Jenkins2009PRB,Jenkins-overdopedPCCO}

%%%%%%%%%%%%%%%%%%%%%%%%%%%%trim=l b r t [scale=.45,clip=true, trim = 50 0 180 15]
\begin{figure*}[!t]
\includegraphics[scale=.5,clip=true]{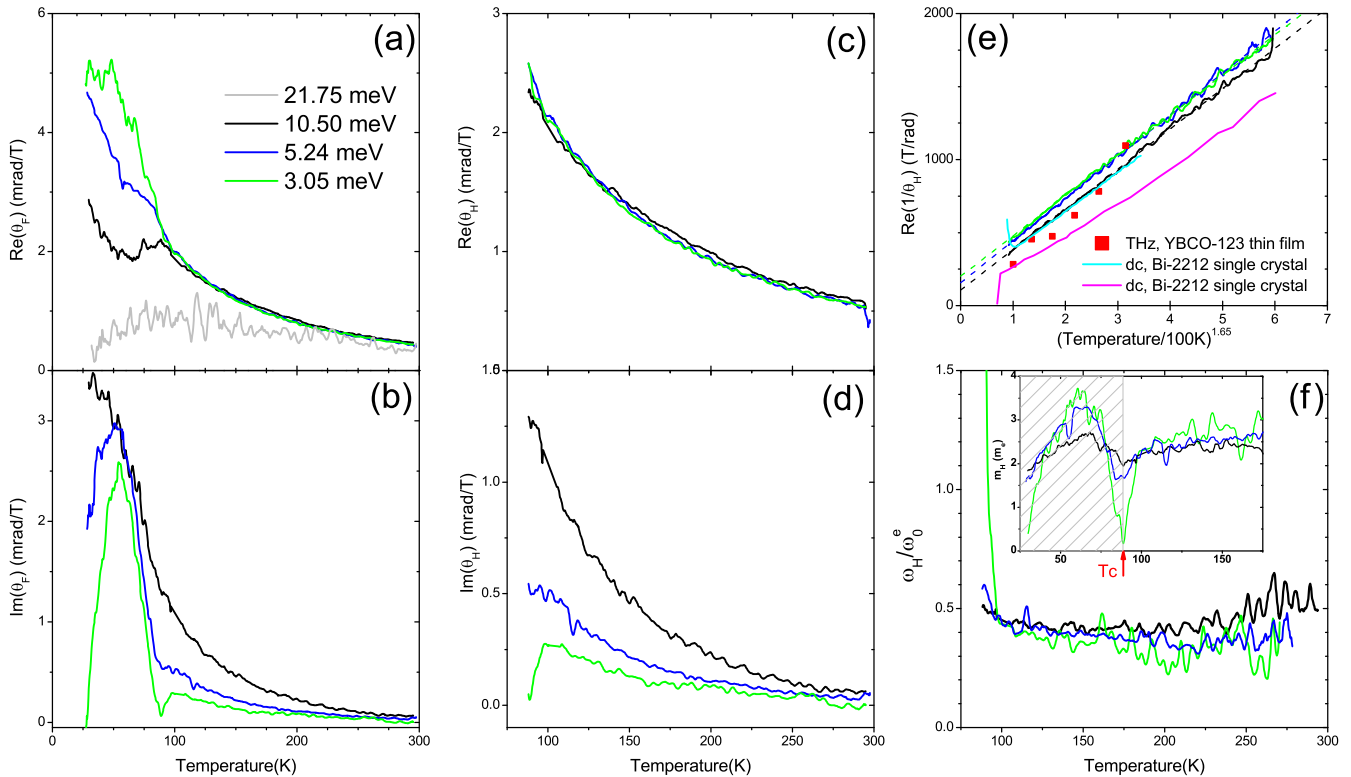}% Here is how to import EPS art
\caption{\label{Fig1}(color online) (a,b) Real and Imaginary parts
of the Faraday angle measured at 3.05, 5.24, 10.5, and 21.75\,meV.
The imaginary part of the Faraday angle for the 21.75\,meV data
set was not reliably measured so is not reported. (c,d) Real and
imaginary part of the Hall angle in the normal state. (e)
$Re(1/\theta_H)$ measurements are consistent with THz measurements
in the range 0.41 to 1.24\,meV performed on optimally doped
YBCO-123 films (red squares) \cite{Parks_THzHall_YBCO123} as well
as dc-$\cot \theta_H$ measurements performed on slightly overdoped
(T$_c$=86.5\,K) (cyan) \cite{Xu_dcHall_overBi2212} and optimally
doped single crystal Bi-2212
(violet).\cite{Forro_dcHall_OptBi2212} The dashed straight lines
are linear fits to the IR Hall data. (f) The measured Hall
frequency normalized to the free electron Hall frequency,
$\omega^e_0 = 0.115\,\text{meV/T}$. The inset shows the Hall mass
in units of bare electron mass $m_e$ versus temperature. The
superconducting state is cross-hatched in grey since the single
fluid Drude parametrization does not apply to the superconducting
state.}
\end{figure*}
%%%%%%%%%%%%%%%%%%%%%%%%%%%

\section{Experimental Description}

Optimally doped single crystal $\rm Bi_2 Sr_2 Ca Cu_2 O_{8+x}$
grown by the travelling floating zone method
\cite{GendaGuFloatingZoneMethod} were exfoliated normal to the
c-axis and mounted to a z-cut quartz substrate  with a broadband
nichrome antireflection
coating.\cite{RSI-Jenkins,GJThesis,SchmadelPRBRapid,SchmadelThesis}
The thickness was determined to be 100\,nm with an area defined by
a 2.5\,mm diameter circular aperture. A midpoint T$_c$ of 87\,K
with a width of 1.5\,K was measured using an ac magnetic
susceptibility probe.

The Faraday rotation and circular dichroism (expressed as the
complex Faraday angle, $\theta_{\rm F}=t_{yx}/t_{xx}$, the ratio
of the off-diagonal and diagonal Fresnel transmission amplitudes)
as well as the relative transmission were measured at a set of
discrete frequencies as a continuous function of temperature. The
output of a far-infrared molecular vapor laser was polarization
modulated via a rotating quartz quarterwave plate and subsequently
transmitted through the c-axis oriented sample at normal incidence
in applied magnetic fields up to 8T. The complex Faraday angle was
obtained by harmonically analyzing the detector signal, a
technique that is detailed elsewhere.\cite{RSI-Jenkins} Both the
real and imaginary parts of the Faraday angle measured at fixed
temperature were linear in applied field. The Faraday angle as a
continuous function of temperature is presented in
Fig.\,\ref{Fig1}(a,b). The imaginary part of the Faraday angle
measured at 21.75\,meV could not reliably be measured due to
particularly low optical throughput power.

In the normal state, the complex Hall angle is derived from the
Faraday angle in the thin film limit via $\theta_{\rm H} = (1+
\frac{n+1}{Z_0 \sigma_{\rm xx} d}) \theta_{\rm F}$, where
$\sigma_{\rm xx}$ is the longitudinal conductivity, $n$ is the
index of refraction of the quartz substrate, $Z_0$ is the
impedance of free space, and $d$ is the thickness of the
film.\cite{RSI-Jenkins} FTIR-spectroscopic transmission
measurements were performed in the spectral range from 2 to
25\,meV at a set of discrete temperatures ranging from 100 to
300\,K.  The complex conductivity $\sigma_{\rm xx}$ was extracted
by fitting to a simple Drude form. $\theta_H$ is very insensitive
to errors in $\sigma_{\rm xx}$ since the conversion factor was
found to be $<15\%$ at all measured frequencies and temperatures.
\cite{GJThesis}

The resulting normal state complex Hall angle measured at 3.05,
5.24, and 10.5\,meV as a function of temperature is shown in
Fig.\,\ref{Fig1}(c,d). Both the real and imaginary parts of the
Hall angle at all temperatures and frequencies are positive
indicating a net holelike Hall response.

\section{Normal state data analysis: Drude Model}

Spectroscopic measurements show the longitudinal conductivity well
described by a simple Drude model in the FIR where the scattering
rate is proportional to temperature and the plasma frequency is
temperature independent \cite{Quijada,Tu_Bi2212} consistent with
early ARPES measurements \cite{Valla_Bi2212_ARPES} and dc
resistivity measurements.\cite{Quijada} Motivated by these
experimental results, we analyze our FIR Hall results in terms of
a simple Drude parametrization. It should be noted, however, that
the well-known anomalous Hall
effect\cite{Konstantinovic_dcHall_Bi2212} where dc-$\cot
{\theta_H} \sim T^{1.78}$ together with the dc resistivity $\sim
T$ is difficult to reconcile within a Drude parametrization. One
of our initial motivations for extending dc Hall angle
measurements to THz frequencies was to separately determine the
temperature dependences of the Hall scattering rate and the Hall
mass.

Within a Drude parameterization, $\tan{\theta_H} \approx \theta_H
= \omega_H/(\gamma_H- i \omega)$ where $\omega_H=e B / c\,m_H$ is
the Hall frequency, $m_H$ is the Hall mass, $\gamma_H$ is the Hall
scattering rate, $\omega$ is the applied optical frequency, B is
the applied magnetic field, c is the speed of light, and e is the
bare charge of an electron. Rearranging, we may write:
\begin{eqnarray}
&Re(1/{\theta_H}) = \gamma_H/\omega_H  \notag\\
&\frac{\omega_H}{\omega_0^e}= \frac{\omega}{\omega_0^e}
Im(1/\theta_H)^{-1} \label{eq:HAdefinition}
\end{eqnarray}
or, equivalently, expressing the Hall frequency in terms of the
Hall mass gives:
\begin{eqnarray}
\frac{m_H}{m_e}  = -\frac{\omega_0^e}{\omega} Im(1/\theta_H)
\end{eqnarray}
where $\cot{\theta_H} \approx 1/\theta_H$, $\omega_0^e = 0.115$
meV/T is the bare electron cyclotron frequency and $m_e$ is the
bare electron mass. Figures \ref{Fig1}(e) and (f) show
$Re(1/{\theta_H})$, $\omega_H$, and $m_H$ at several FIR
frequencies as a function of temperature.

$\omega_H$ (and $m_H$) is frequency independent (within the errors
of the measurement) and only slightly temperature dependent.  As
temperature increases or frequency decreases, the $\omega_H$ data
becomes more noisy commensurate with the decreasing magnitude of
the Hall angle. The best signal-to-noise for determining the
temperature dependence of $\omega_H$ are the 10.5\,meV and
5.24\,meV data sets which show a very discernable bowed behavior.

Even at finite frequencies up to 10.5\,meV, $Re(1/{\theta_H})$
obeys a power law $\sim T^{1.65\pm 0.1}$ consistent with published
THz Hall measurements performed on optimally doped YBCO-123 films
at 0.41\,meV to 1.24\,meV \cite{Parks_THzHall_YBCO123} and dc
measurements performed on Bi-2212 single crystals
\cite{Xu_dcHall_overBi2212,Forro_dcHall_OptBi2212} and
films.\cite{Konstantinovic_dcHall_Bi2212} Although the temperature
power laws are the same, varying offsets between data sets are
presumably from variations in sample quality and associated
impurity scattering. A small frequency dependence is discernable
which shows $Re(1/{\theta_H})$ decreasing slightly with increasing
frequency consistent with earlier FIR broadband measurements on
optimally doped YBCO-123 near $T_c$. \cite{Grason}

The Drude analysis is easy to understand but very limited. It
assumes that two parameters, a characteristic frequency and
scattering rate, can reasonably capture the physics of all
transport measurements. However, Fermi surfaces which have
anisotropic scattering rates and Fermi velocities will manifest as
differing effective Drude parameters when probed with various
experimental techniques due to the nature of averaging around the
FS. In order to compare the IR Hall data with ARPES and other
transport data, we review Boltzmann formalism.

\section{Boltzmann Theory: Transport and ARPES data comparisons}

Within the RTA in Boltzmann theory, the off-diagonal and diagonal
conductivities may be expressed as path integrals around the FS
involving the Fermi velocity, scattering rate, and size and shape
of the FS:\cite{DrewMillisSigmaXX}
\begin{align}\label{eq:RTA}
\sigma_{xy}&=\frac{e^2}{h c_0}\frac{e \vec{B}}{\hbar c}  \cdot
\oint_{\text{\tiny{FS}}} dk \frac{\vec{v}_k^* \times d\vec{v}_k^*/dk}{(\gamma^*_k - i \omega)^2} \notag\\
\sigma_{xx}&=\frac{e^2}{h c_0} \oint_{\text{\tiny{FS}}} dk
\frac{|\vec{v}_k^*|}{\gamma^*_k - i \omega}
\end{align}
where  $v^*_k$ is the (renormalized) Fermi velocities as measured
by ARPES, $\gamma^*_k= v^*_k \, \Delta k$, $\Delta k$ is the
momentum distribution curve (MDC) width as measured by ARPES, and
$c_0$ is the average interplane spacing. Various transport
quantities can be calculated, but the relevant quantities of
interest are the dc Hall coefficient
$R_H=\sigma_{xy}/\sigma_{xx}^2$, the Hall angle
$\theta_H=\sigma_{xy}/\sigma_{xx}$, and the Hall frequency and
scattering rate given by the expression $\omega_H/(\gamma_H-i
\omega) = \sigma_{xy}/\sigma_{xx}$.

In a comparison of ARPES and transport data it is important to
recognize that the current relaxation time that characterizes
transport is not necessarily the same as the quasiparticle
lifetime measured by ARPES.  For example, small angle elastic
scattering does not affect transport but does contribute to MDC
width broadening.\cite{DrewMillisSigmaXX}  However, it is equally
important to realize that for the rather simple Fermi surfaces
under consideration in this study, $R_H$ and $\omega_H$ (and
$m_H$) depend only weakly on the anisotropy of the scattering
rate. In the limit of high frequency or isotropic scattering,
$R_H$ \cite{Ong_FSIntegrals} and $\omega_H$ are independent of
scattering rate, and are given by:
\begin{align}\label{eq:isoscattering}
\omega_{H}&=\frac{e \vec{B}}{\hbar c}  \cdot
\frac{\oint_{\text{\tiny{FS}}} \vec{v}_k^* \times d\vec{v}_k^*}{\oint_{\text{\tiny{FS}}} dk |\vec{v}_k^*|} \notag \\
R_H&=\frac{h c_0}{e^2} \frac{e \vec{B}}{\hbar c} \cdot
\frac{\oint_{\text{\tiny{FS}}} \vec{v}_k^* \times
d\vec{v}_k^*}{(\oint_{\text{\tiny{FS}}} dk |\vec{v}_k^*|)^2}
\end{align}
Note that for the case of isotropic mean free path,
$v_k^*/\gamma^*_k$, and a single sign of concavity associated with
the entire FS, the dc Hall coefficient is given by $R_H= R_H^{Lut}
(C/C')^2$ where $C$ and $C'$ are the circumferences of a circular
and non-circular Fermi surface of the same area, and $R_H^{Lut}$
is the Luttinger value of the Hall coefficient.

The cyclotron frequency $\omega_c$ (or cyclotron mass $m_c$) is
inherently independent of scattering rate and depends only on the
Fermi velocity and circumference of the FS:
\begin{align}\label{eq:omegac}
\frac {\omega_c}{\omega_0^e}= (\frac {m_c}{m_ e})^{-1}= m_e
\frac{e B}{\hbar c} \frac{2 \pi}{\oint_{\text{\tiny{FS}}}
\frac{dk}{|\vec{v}_k^*|}}
\end{align}
where $\omega_0^e$ is the free electron cyclotron frequency and
$m_e$ is the bare electron mass. The Hall frequency is exactly
equal to the cyclotron frequency in the special case of a circular
FS with isotropic velocity and isotropic scattering.

\section{Substantiating comparisons between ARPES+RTA and transport In Cuprate Systems}

ARPES measurements reflect the bulk band dispersion and FS
topology in various cuprate systems. In overdoped PCCO in the
T$\to$0 limit, the measured dc-$R_H$ agrees with the value
calculated from ARPES data within the RTA, and both are consistent
with Luttinger's thereom.\cite{Jenkins-overdopedPCCO} In
underdoped n-type cuprates, the number density of the electron
pocket derived from ARPES measurements is consistent with the
stoichiometric doping as measured in
Sm$_{1.86}$Ce$_{0.14}$CuO$_{4}$\cite{ARPES_edoped_Park} and
Nd$_{2-x}$Ce$_x$CuO$_{4\pm\delta}$
(NCCO).\cite{Jenkins2009PRB,ARPES_edoped_Matsui,Lin_Millis} In
optimally doped Bi-2212, the band dispersion as measured by recent
low energy laser ARPES agrees with earlier higher energy
measurements verifying that both techniques are measuring bulk
properties.\cite{Koralek_LaserARPES}

Before analyzing Bi-2212, it is instructive to compare ARPES and
transport data in strongly overdoped Tl-2201 where deviations away
from simple Fermi liquid-like behavior are much less severe.  The
ARPES measured FS size from reported data is reasonably consistent
with the de Haas van Alphen measurements,\cite{Vignolle_QO_Tl2201}
AMRO measurements,\cite{Peets_AMRO} Luttinger's
thereom,\cite{Damascelli_Tl2201_ARPES,Hussey_Nature2003} and low
temperature Hall coefficient measurements.\cite{mackenzie_Tl_RH}
There is general agreement between ARPES and band structure
calculations,\cite{Peets_AMRO,Singh_DFTBandCalcs_Tl2201} and the
Wiedemann-Franz law has been experimentally
verified.\cite{Proust_WeidermanFranz_Tl2201Overdoped,Peets_AMRO}

These results make a compelling case for the applicability of
Fermi liquid theory in strongly overdoped p-type cuprates.  We
directly test the relaxation time approximation by using Boltzmann
transport theory to calculate dc-$R_H$ (Eq.\,\ref{eq:RTA}) and
$m_c$ (Eq.\,\ref{eq:omegac}) with parameters derived from ARPES
data. The size and shape of the FS are given by the following
tight binding parameters: $\mu=0.2438$, $t_1 =-0.725$, $t_2
=0.302$, $t_3 =0.0159$, $t_4 = -0.0805$, $t_5 = 0.0034$, all
expressed in eV.\cite{Damascelli_Tl2201_ARPES} The scattering rate
as a function of angle around the FS is
reported.\cite{Damascelli_Tl2201_ARPES} The nodal and antinodal
Fermi velocities are given by $v_{\pi,\pi}=2.0 \pm 0.15
\text{\,eV\,\AA}$   and $ v_{\pi,0}=0.765 \text{\,eV\,\AA}$,
respectively. \cite{privatecomm_DamascelliandGiorgio_Tl2201}
Interpolation between the nodal and antinodal Fermi velocities is
reasonably provided by the tight binding model. However, to
conservatively estimate errors in the interpolation, a range of
functional forms of the Fermi velocity given by
$v_k=v_{\pi,\pi}+(v_{\pi,0}-v_{\pi,\pi})(\sin{2 \theta})^n$ are
used where n ranges from 2 to 8 and the location on the FS is
parameterized by the angle $\theta$.

The integrated area of the ARPES measured FS yields a number
density consistent with the Luttinger value associated with the
stoichiometric doping, 1.26. The Hall coefficient calculated from
the stoichiometric number density yields $R_H=8.4 \times 10^{-10}
\text{\,m$^3$/C}$, consistent with the low temperature measurement
of dc-$R_H=8.5 \times
10^{-10}\text{\,m$^3$/C}$.\cite{mackenzie_Tl_RH} However, the
measured dc-$R_H$ increases to a peak value of about $12 \times
10^{-10} \text{\,m$^3$/C}$ at $\sim$100\,K and, importantly, falls
off to a value of about $10 \times 10^{-10}\text{\,m$^3$/C}$ at
higher temperatures $\sim$300\,K.

Evaluating Eq.\,\ref{eq:RTA} yields a Hall coefficient in the
range of $7.5$ to $10 \times 10^{-10}\text{\,m$^3$/C}$. The range
of values results from the various functional forms assumed for
the interpolated values of $v^*_k$ as well as some errors
introduced from uncertainties associated with the ARPES measured
scattering rates. Good agreement between the ARPES+RTA value and
the low temperature dc-$R_H$ value is achieved for the case where
$n \approx 3$ for the velocity anisotropy interpolation function.

The measured dc-$R_H$ value at 100\,K lies well outside the range
that is predicted by ARPES (assuming that the ARPES measured
velocity and scattering rate anisotropy do not change
significantly between T$_c$ and 100\,K) illustrating that a
breakdown of the simple relaxation time approximation occurs for
Tl-2201 at a doping of 1.26, even considering the conservatively
large error bars. Therefore overdoped Tl-2201 appears to have an
anomalous Hall effect similar to optimally doped p-type cuprates
although it is much weaker.

At sufficiently high temperature, the scattering becomes dominated
by phonon scattering and is expected to become isotropic. The Hall
coefficient is then independent of scattering rate (see
Eq.\,\ref{eq:isoscattering}). Under the assumption of isotropic
scattering and Fermi velocities which agree with ARPES
measurements, the predicted ARPES value of the Hall coefficient is
$9.5\pm0.5 \times 10^{-10}\text{\,m$^3$/C}$  in reasonable
agreement with the dc-$R_H$ value of about $10 \times
10^{-10}\text{\,m$^3$/C}$ at 300\,K. Whatever the cause of the
anomalous dc Hall effect in overdoped Tl-2201, the effect seems to
disappear at low temperature and in the vicinity of room
temperature. At very high temperature in other cuprate systems,
the Hall coefficient continues to decrease due to thermally
activated transitions to other
bands.\cite{Ando_dcRH_LSCOvsHighT,BasovMIRCuprates}

For completeness, it should be noted that under the assumptions of
isotropic mean free path,  the Hall coefficient found from
integrating the nearly circular FS yields the Luttinger value as
expected which is consistent with the dc measured value at low
temperature. However, the ARPES data shows that neither isotropic
scattering rate nor isotropic mean free path occurs at low
temperature.

The cyclotron mass $m_c$ calculated from Eq.\,\ref{eq:omegac}
ranges from $4.5\,m_e$ to $5.7\,m_e$, a result consistent with
quantum oscillations measurements where
$m_c=4.1\pm1.0$.\cite{Vignolle_QO_Tl2201} The associated error
bars are large for both the ARPES+RTA value as well as the
measured value. If a more accurate $m_c$ is measured by quantum
oscillation experiments and more ARPES velocities are measured so
as to derive the functional form of the velocity around the FS, a
more thorough comparison could be made.

The Tl-2201 case substantiates the ARPES+RTA methodology for the
hole doped cuprate systems while concurrently showing an
interesting beakdown of the RTA in the vicinity of 100\,K, even in
the strongly overdoped regime. As we show in the next section, the
discrepancy between dc-$R_H$ and the ARPES+RTA value is much
larger in optimally doped Bi-2212 than for strongly overdoped
Tl-2201, but both show an enhanced Hall response above the
ARPES+RTA value. Importantly, the IR Hall response also shows an
enhancement above the ARPES+RTA value.

\section{Bi-2212: Hall data comparison with ARPES+RTA}\label{Bi2212-ARPESComparison}

There are complications which arise when comparing Bi-2212 ARPES
results to transport data. The nodal region is very well
characterized, but intracell bilayer coupling  complicates the
analysis in the vicinity of the antinode. Recent ARPES
measurements have resolved the bilayer splitting in optimally
doped Bi-2212.\cite{Chuang_BilayerSplittingEnergy_Bi2212,
Feng_Bilayersplitting}  Early measurements reported an anisotropy
in the scattering rate of 2 to 3 between the nodal and antinodal
points,\cite{Valla_Bi2212_ARPES} but they were not resolving the
bilayer splitting. It has been well argued that these early ARPES
measurements on Bi-2212 grossly overestimated the scattering rate
away from the nodal direction due to the unresolved bilayer
splitting.\cite{Kordyuk_ARPES_Bi2212,Damascelli_RMP_ARPESReview}

However, as we will demonstrate, the dc and FIR Hall response is
much larger than the ARPES+RTA prediction. Effects produced by
bilayer splitting and anisotropic scattering are small compared to
this discrepancy, and both effects when fully taken into account
tend to \textit{increase} the discrepancy.

We begin our analysis by assuming an isotropic scattering rate,
but consider anisotropic scattering effects later. An isotropic
Fermi velocity with a large holelike FS was measured by
ARPES.\cite{Valla_Bi2212_ARPES, Kaminski_optBi2212_ARPES} The
Fermi velocity was measured to be about $1.8 \text{\,eV\,\AA}$ at
100\,K.\cite{Valla_Bi2212_ARPES, Plumb_LaserARPES_vFNodevsT,
Yamasaki_Bi2212LaserARPES_NodalScatteringRate} The shape of the FS
is given by a tight binding model where the dispersion is given by
$\epsilon_p = - 2 \, t_1\, ( \cos{p_x} + \cos{p_y}) + 4 \, t_2 \,
\cos{p_x}\cos{p_y} - 2 \, t_3 \, ( \cos{2 p_x} + \cos{2 p_y})$
with hopping parameters $t_1=0.38$ eV, $t_2=0.32 t_1$, and
$t_3=0.5 t_2$ \cite{
LDA_BandStructure_Calculation_Overdoped_edoped, Lin_Millis,
Millis_tightbindingparameters} and a chemical potential chosen
such that the area of the FS is commensurate with Luttinger's
thereom. This model was modified to include bilayer splitting
consistent with ARPES measurements on optimally doped Bi-2212 by
adding a gap function given by $\Delta_0 (\cos{p_x x} + \cos{ p_y
y})^2$ where $\Delta_0 \approx \pm
50$meV.\cite{Chuang_BilayerSplittingEnergy_Bi2212,
Feng_Bilayersplitting} Under the assumptions of isotropic
scattering and isotropic Fermi velocity (assuming the ARPES value
of $1.8 \text{\,eV\,\AA}$) for the case where bilayer splitting is
ignored (setting $\Delta_0=0$), we find $\omega_{H0} = \omega_{c0}
= 0.33 \, \omega_0^e$ ($m_{H0} =3.0\, m_e$) where
$\omega_0^e=0.115$\,meV/T is the free electron cyclotron
frequency, and dc-$R_{H0}=5.8 \times 10^{-10}\text{\,m$^3$/C}$.
The deviation from the Luttinger value of dc-$R_H=6.0 \times
10^{-10}\text{\,m$^3$/C}$ is due to the slightly non-circular FS.

If the velocity anisotropy given by the tight binding dispersion
for the $\Delta_0=0$ case are assumed (where the nodal velocity is
scaled to the ARPES measured value of $1.8 \text{\,eV\,\AA}$),
then it is found that deviations from the above calculated
transport values are very small. The reason is that the tight
binding model shows only a slight anisotropy of the Fermi velocity
where the nodal and antinodal velocity are approximately equal
with a slight depression of about 15\% in between.

When we include the bilayer splitting where $\Delta= +50$ meV
($-50$ meV) which changes the dispersion away from the nodal point
(causing larger anisotropic variation of the Fermi velocity) as
well as the size and shape of the Fermi surface, we find a
suppression of the Hall frequency $\omega_H/\omega_0^e=0.29$
($0.32$), a cyclotron frequency of $\omega_c/\omega_0^e = 0.25$
($0.39$), and a Hall coefficient of $R_H = 5.1\times
10^{-10}\text{\,m$^3$/C}$ ($6.0\times 10^{-10}\text{\,m$^3$/C}$).
The Boltzman transport values predicted by ARPES including bilayer
splitting are then $R'_{H0}=5.5 \times 10^{-10}\text{\,m$^3$/C}$,
$\omega'_{c0}=0.25\, \omega_0^e$ and $0.39\, \omega_0^e$ for the
two Fermi surfaces, and $\omega'_{H0} = 0.30 \, \omega_0^e$
(corresponding to $m'_{H0} = 3.3 \, m_e$).

We introduce a scattering rate anisotropy of the form
$\gamma(\theta) = 2-\sin(2 \theta)$ to the same tight binding
model such that the antinodal point scattering is twice as large
as the nodal scattering, an overestimate of the anisotropy. For
the $\Delta_0=0$ case, we find $R_{H0}=7.8 \times
10^{-10}\text{\,m$^3$/C}$, and $\omega_{H0}$ values ranging from
$0.34$ to $.33 \, \omega_0^e$ corresponding to the low and high
frequency limits, respectively. If the bilayer splitting is then
included while concurrently including the anisotropic scattering
which is assumed to be the same in both the antibonding and
bonding bands, then the values become $R'_H= 7.4 \times
10^{-10}\text{\,m$^3$/C}$ and $\omega'_H$ ranges from $0.32$ to
$.30 \, \omega_0^e$ between the low and high frequency limits,
respectively.

Summarizing, using ARPES data for optimally doped
Bi-2212,\cite{Valla_Bi2212_ARPES, Kaminski_optBi2212_ARPES,
Chuang_BilayerSplittingEnergy_Bi2212,
Yamasaki_Bi2212LaserARPES_NodalScatteringRate,
Plumb_LaserARPES_vFNodevsT, Feng_Bilayersplitting,
Damascelli_RMP_ARPESReview, Koralek_LaserARPES,
privatecomm_Shens_student,Kordyuk_ARPES_Bi2212} $R_H$, $\omega_c$,
and $\omega_H$ (and $m_H$) were calculated within Boltzmann
transport formalism. The ARPES+RTA value of $\omega_H =
0.30\,\omega_0^e$ ($m_H = 3.3 m_e$) is substantially less than the
measured value of $\omega_H = 0.44 \pm .04\,\omega_c^e$ at 100\,K
for all frequencies $\leq 10.5$\,meV. This represents a
substantial enhancement of the FIR Hall response $\sim 50\%$ above
the ARPES+RTA expectation. Similarly, the measured dc-$R_H \approx
20 \times 10^{-10}\text{\,m$^3$/C}$ at 100\,K
\cite{Konstantinovic_dcHall_Bi2212, Ri_dcHall_OptBi2212,
Forro_dcHall_OptBi2212, Xu_dcHall_overBi2212} is enhanced by a
factor of 3 to 4 above the ARPES+RTA calculated value of $R_{H} =
5.5 \times 10^{-10} \text{\,m$^3$/C}$.

%%%%%%%%%%%%%%%%%%%%%%%%%%%%trim=l b r t [scale=.45,clip=true, trim = 50 0 180 15,, trim = 85 350 30 60]
\begin{figure*}[!t]
\includegraphics[scale=.27,clip=true,trim = 25 0 320 0]{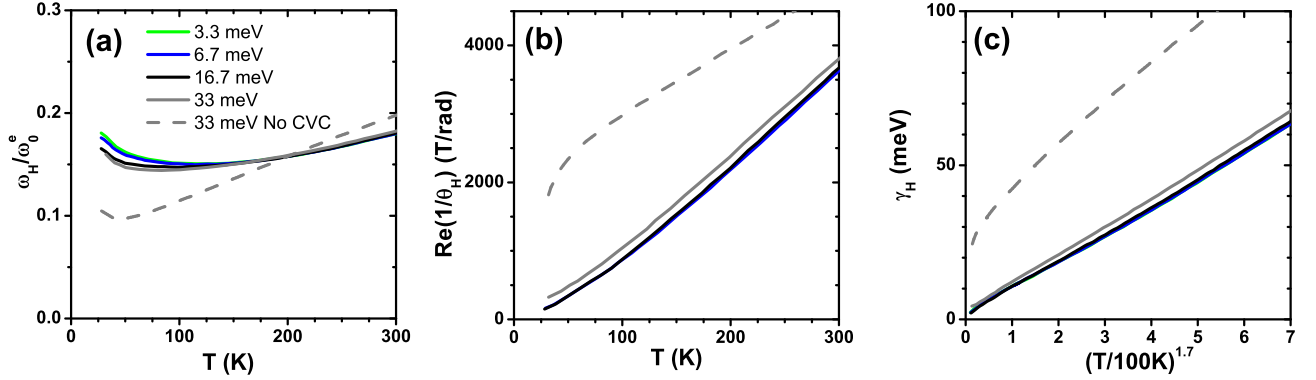}
\caption{\label{Fig2}(color online) The calculated Hall angle
using a FLEX + T-matrix approximation parameterized in the same
manner as the reported data in Fig.\,\ref{Fig1} for easy
comparison. Shown are graphs of the (a) Hall frequency $\omega_H$
normalized to the free electron Hall frequency,
$\omega_0^e=0.115$\,meV/T, (b) $Re(1/ \theta_H)$, and (c) Hall
scattering rate $\gamma_H$. The green (3.3 meV), blue (6.7 meV),
and black (16.7 meV) plots in figures (b) and (c) are nearly
indistinguishable.}
\end{figure*}
%%%%%%%%%%%%%%%%%%%%%%%%%%%

%%%%%%%%%%%%%%%%%%%%%%%%%%%%%%%%%%%%%%%%%%%%%%%%%%%%%%%%%%%%%%%%%
\section{Comparisons with FLEX+CVC model}

A theory of cuprate transport properties which includes
electron-electron interactions mediated by antiferromagnetic
fluctuations has been developed and extensively applied by
Kontani.\cite{Kontani_2008Review} Antiferromagnetic fluctuations
were calculated within the FLEX approximation, and the effective
low energy electron-electron interaction Hamiltonian is then
proportional to the dynamical spin susceptibility. Electron-magnon
scattering causes corrections to the self energy as well as
current vertex corrections (CVCs) to the conductivity. CVCs change
the magnitude of the k-dependent current and cause deviations of
the current direction away from the FS normal. A T-matrix
formalism has also been incorporated to account for
superconducting fluctuation effects in the vicinity of $T_c$.

This theory has proven highly successful in accounting for the
features of the anomalous behavior of the dc Hall effect in both
electron and hole doped cuprates as a function of doping and
temperature.  More recently it has been applied to the frequency
dependence of hole doped cuprates at mid-IR frequencies and
electron doped cuprates at THz frequencies.  The theory
successfully accounted for the doping, temperature, and frequency
dependence of the IR Hall data on overdoped electron doped
cuprates.\cite{Jenkins-overdopedPCCO,Jenkins2009PRB}

Calculated results of the IR Hall response are shown in
Fig.\;\ref{Fig2}. The calculation considers a one band Hubbard
model for the copper-oxygen planes with tight binding parameters
for YBCO-123 (which is similar to Bi-2212 except for the presence
of the chain bands) with a Coulomb interaction characterized by an
on site potential U. The hole doping is set by the chemical
potential. The longitudinal and Hall conductivity are calculated
as a function of temperature for several representative THz
frequencies.  The calculations are performed both including and
ignoring CVCs.  The later case is equivalent to the RTA within the
FLEX approximation. To compare with the experimental data,
Re($1/\theta_H$), $\omega_H$, and $\gamma_H$ are calculated from
the conductivity tensor via the definition of the Hall angle and
Eq.\,\ref{eq:HAdefinition}.

The qualitative behavior of $\omega_H$ and Re(1/$\theta_H$) is
consistent with the IR Hall data. Both $\omega_H$ and
Re(1/$\theta_H$) in Fig.\,\ref{Fig2}(a,b) exhibit very little
frequency dependence consistent with the data in
Fig.\,\ref{Fig1}(e,f). The slight upward bowing of $\omega_H$ with
temperature and the superlinearity of Re(1/$\theta_H$) is
reproduced (although the power law is somewhat weaker).

Previously reported calculations of dc-$R_H$ in
Ref.\,\onlinecite{Kontani_2008Review}, Fig.\,19(i,iii), show a
very large enhancement of dc-$R_H$ above that predicted by the RTA
(the `No CVC' case) consistent with the experimentally measured
enhancements in optimally doped Bi-2212 above the ARPES+RTA
predicted value calculated in the previous
section.\cite{Konstantinovic_dcHall_Bi2212}

Similar to the dc-$R_H$ case, the enhancements of the IR Hall
response $\omega_H$ above that expected from the ARPES+RTA value
is also quantitatively equal to the enhancements calculated by
including CVCs. As shown in Fig.\,\ref{Fig2}(a), the enhancement
of $\omega_H$ by the inclusion of CVCs at 100\,K and 33\,meV is
$\omega_H^{CVC}/\omega_H^{NoCVC}=1.3$. This is consistent with the
enhancement of the experimental value at 100\,K and 10.5\,meV
above the ARPES+RTA value,
$\omega_H^{FIR}/\omega_H^{ARPES+RTA}=1.5$ where
$\omega_H^{FIR}=0.44 \, \omega_0^e$ from Fig.\,\ref{Fig1}(f),
$\omega_H^{ARPES+RTA} \approx 0.30 \, \omega_0^e$ (which includes
bilayer splitting effects), and $\omega_0^e=0.115$\,meV/T is the
free electron cyclotron frequency. The smaller enhancement of
$\omega_H$ at THz frequencies compared with that of dc-$R_H$ is
well reproduced by the theory.

The calculations capture all of the qualitative features of the
data. This is an impressive success of the theory since the
frequency and temperature dependence of the Hall data, as well as
the quantitative discrepancy with the ARPES+RTA value, places
severe constraints on theories.

Although the similarities between the data of
Fig.\,\ref{Fig1}(e,f) and the calculated values of
Fig.\,\ref{Fig2}(a,b) are striking, there are some important
discrepancies. In particular, the magnitude of $\omega_H$ differs
from the experimental value.

To address this issue, we note that the FS resulting from the
model calculation differs significantly from the FS as measured by
ARPES in Bi-2212. The tight binding model parameters are $t_1 =
0.35$\,eV, $t_2 = 0.17 t_1$, and $t_3 =0.2 t_1$. The chemical
potential is set to achieve the correct FS area corresponding to
the stoichiometric doping. For U=0 the FS is very similar to that
measured by ARPES.  However, for nonzero U the Fermi surface
becomes distorted due to self energy corrections.  The FS
distortions result in much more curvature of the Fermi surface in
the antinodal regions and less curvature in the vicinity of the
nodal regions (compare the measured FS from
Ref.\,\onlinecite{Feng_Bilayersplitting} and
\onlinecite{Chuang_BilayerSplittingEnergy_Bi2212} with Fig.\,2(a)
of Ref.\,\onlinecite{Kontani_1999}). Since $\omega_H$ weights more
heavily the higher curvature regions of the FS, which for
Kontani's FS occurs near the lower velocity sections near the van
Hove singularity in the vicinity of the antinode, the calculated
value of $\omega_H^{NoCVC}$ is expected to be substantially lower
than our calculated value $\omega_H^{ARPES+RTA}$.

In view of these FS differences between the model calculation and
that observed by ARPES, it is not surprising that the model
$\omega_H^{NoCVC} \approx .11\, \omega_0^e$ is about a factor of 3
smaller than the ARPES+RTA derived value of $\omega_H^{ARPES+RTA}
\approx 0.30\, \omega_0^e$ although these values, in principle,
should match.

Similarly, the experimental value of
$\omega_H^{FIR}=0.44\,\omega_0^e $, is roughly a factor of 3
larger than the model value of $\omega_H^{CVC} = 0.15\,\omega_0^e
$. Again the magnitude is off but, most importantly, the
enhancement ratio is approximately correct  such that
$\omega_H^{CVC}/\omega_H^{NoCVC} \sim
\omega_H^{FIR}/\omega_H^{ARPES+RTA}$.

These absolute magnitude comparisons are exacerbated by another
factor not included in the conductivity calculations.  Mott
correlations associated with the Coulomb repulsion are expected to
shift conductivity spectral weight from the low energy Drude-like
response to high energies
$\omega>$U.\cite{Millis_tightbindingparameters}  Moreover, the
spectral weight associated with $\sigma_{xy}$ is suppressed by a
factor of $\sim$3 more than $\sigma_{xx}$ as demonstrated by an
analysis of the FIR and MIR Faraday measurements on
Bi-2212.\cite{SchmadelPRBRapid} Since these Mott correlation
effects are not included in the model, one expects an additional
level of over estimation of the calculated $\omega_H$.

The results on Bi-2212 should be contrasted to our previous
results in overdoped PCCO where the Hall frequency calculated from
the FLEX+CVC theory is larger than our measured value as one would
expect from the omitted effects of spectral weight renormalization
due to Mott correlations from the theory.
\cite{Jenkins-overdopedPCCO} The effects from FS distortions due
to the Hubbard U are smaller for n-type cuprates as can be seen in
Fig.\,2 of Ref.\,\onlinecite{Kontani_1999}. Since the FS is
smaller, the portion near ($\pi$,0) is further from the magnetic
Brillouin zone and the van Hove singularity. Therefore, reductions
of the Fermi velocity due to the addition of the Hubbard U to the
tight binding model are much less severe resulting in less
reduction of $\omega_H$ calculated within the RTA approximation.

The good agreement of the FLEX+CVC theory in accounting for the
enhancements of the IR Hall response above that predicted by the
ARPES+RTA value while reproducing the frequency and temperature
dependence is significant evidence that current vertex corrections
to the conductivity together with interactions mediated by
magnetic fluctuations are the origin of the anomalous Hall effect
in the cuprates, an important conclusion of this work.  However,
the task of removing some deficiencies of the theory remain such
as incorporating a proper treatment of Mott correlations on the
conductivity spectral weight and self consistently determining the
Coulomb U to produce the measured ARPES FS.

\section{Precursive superconductivity}

In the inset of Fig.\,\ref{Fig1}(f), a dip feature observed in
$m_H$ at 3.05 and 5.24\,meV has an onset temperature well above
$T_c$, but is most pronounced at 3.05\,meV. This appears to be
precursive superconducting behavior well above the transition
temperature. Various experiments have exhibited precursive
superconducting behavior in the normal state in the high-Tc
cuprates: microwave cavity,\cite{BonnMicrowavePrecursor} Nernst
effect,\cite{OngNernstPrecursor} specific
heat,\cite{LoramSpecificHeatPrecursor} infra-red
reflectivity,\cite{MolegraafIRSpecPrecursor} dc- Hall
angle,\cite{MattheyDCHallPrecursor,
BucherDCHallPrecursor,JinDCHallPrecursor} scanning tunnelling
microscopy,\cite{RennerTunnelingPrecursor} and ARPES
measurements.\cite{CampuzanoAREPSPrecursor}

Of particular interest, Corson \textit{et al.} observed a similar
phenomenon above T$_c$ in both under-\cite{SCPrecursorFIR} and
optimally-\cite{SCPrecursorFIRunderdoped} doped Bi-2212 in the
longitudinal conductivity $ {\sigma}_{xx}$ using a terahertz
spectroscopy technique. The real and imaginary part of
$\sigma_{xx}$ of underdoped Bi-2212 (T$_c$=75\,K) as a function of
temperature at 100\,GHz $\sim 0.14$\,meV reveal that the ratio of
the phase of the conductivity
$\phi_{xx}=Im(\sigma_{xx})/Re(\sigma_{xx})$ at T$_c$ to
$\phi_{xx}$ at 100\,K is $\gtrsim 4$ indicating the imaginary part
of $\sigma_{xx}$ increases much more rapidly than the real part of
$\sigma_{xx}$ upon decreasing temperature towards
T$_c$.\cite{SCPrecursorFIRunderdoped}   On optimally doped
Bi-2212, only the real part of $\sigma_{xx}$ as a function of
temperature is reported where deviations away from the Drude
(quasi-particle) contribution measured at frequencies 0.2, 0.6,
and 0.8\,THz is observed.\cite{SCPrecursorFIR}  The magnitude of
the conductivity enhancement above the Drude contribution is a
factor of 2, 1.2, and 1.13, respectively, when temperature is
reduced from 100\,K to T$_c$.

From our transmission measurements as a continuous function of
temperature analyzed in terms of a simple Drude response (reported
in Ref.\,\onlinecite{GJThesis}), we find an enhancement in the
magnitude of the conductivity of 1.3 at $3.05\,\text{meV}\approx
0.7$\,THz and 1.1 at $5.24\,\text{meV} \approx 1.3$\,THz when the
temperature is reduced from 100\,K to T$_c$. The enhancement of
the magnitude of the conductivity is expected to be larger than
the enhancement of only the real part of the conductivity as
measured by Corson \textit{et al.} on optimally doped Bi-2212 due
to the onset of a substantial imaginary part.

The precursive behavior observed in the IR Hall angle is
definitively associated with the longitudinal channel and could
possibly account for the entire behavior. Whether the off-axis
conductivity contributes to the precursive behavior is not
currently discernable.

\section{Superconducting state}
%%%%%%%%%%%%%%%%%%%%%%%%%%%%%%%%%%%%%%%%%%%%%%%%%%%%%%%%%%%%%%%%%%%%%
To analyze the superconducting state, we use a variation of a
phenomenological model of the magneto-conductivity which well
described the FIR low temperature ($<\,20$\,K) magneto-optical
response of superconducting YBCO thin films.
\cite{Kaplanvortexdynamics,
Karaivortexdynamics,ShanLuThesis,SteveLihnthesis} The model
conductivity used to describe the low temperature magneto-optical
data consists of the sum of four temperature independent
Lorentzian oscillators: two associated with vortex core
excitations, one associated with the zero frequency resonance of
the superfluid, and one associated with the thermally excited
nodal quasi-particles.

Unlike the previously reported data, the present study covers a
wide range of temperature. However, it remains instructive to
extrapolate this low temperature model to higher temperatures. The
temperature dependence of the nodal quasi-particle fraction and
scattering rate have been measured in microwave experiments
\cite{Microwave-nsAndQPscatteringvsTemp} whose functional form is
incorporated into the model. All other parameters are assumed
temperature independent.

We note that many parameters can not realistically be assumed
temperature independent. For example, the size of the vortex cores
increase with temperature
\cite{VortexCoreSizeChangeswithTemperature} which decrease the
core level spacing. The pinning frequency depends strongly upon
temperature (for example, it is \textit{a priori} known that the
pinning force vanishes above the vortex glass melting temperature,
$\sim$70\,K).

Regardless, the coarse features of the FIR Hall data are
reproduced (although significant deviations which are discussed in
detail in Ref.\,\onlinecite{GJThesis} notably exist) by the
following minimum number of terms in the conductivity expressed in
the circular basis: a zero frequency superfluid resonance (London
term), a cyclotron resonance associated with the nodal
quasi-particles, and one finite frequency oscillator:
\begin{equation}\label{eq; GJSCmodel}
\begin{split}
\sigma_{\pm} &  = \Big(\frac{f_L}{-\mathit{i} \omega} +
\frac{f_p}{-\mathit{i} (\omega \pm \omega_p) + \Gamma_p (t) }\Big) \notag \\
 & \times ( 1 - f_n (t)\,)
 + \frac{f_n (t)}{-\mathit{i} (\omega \pm
\omega_n) + \Gamma_n}
\end{split}
\end{equation}
where $t=T/T_c$ is the normalized temperature, the nodal
quasiparticle fraction and scattering rate are $f_n (t) =1$ and
$\Gamma_n (t) = 16.3 \; \text{meV} \; t^{1.65}$ for $t>1$, and
$f_n (t) = t^2$ and $\Gamma_n (t) = 16.3 \; \text{meV} \; t^{4}$
for $t<1$, the cyclotron frequency of the nodal quasiparticles $
\omega_n = 0.38 \; \text{meV}$,  the fraction associated with the
number of particles condensed in the superfluid $f_L = .45$, and
the required finite frequency oscillator parameters given by $f_p
= .04$, $ \omega_p = 4.4 \; \text{meV}$, and $\Gamma_p = 2.5 \;
\text{meV}$.

The London term alone in the conductivity causes no Hall effect.
The high frequency 21.75\,meV data is well described by the
quasiparticle cyclotron resonance term independent of other
possible resonant terms, an observation which is consistent with
previous measurements on optimally doped
YBCO.\cite{Karaivortexdynamics} The low frequency data is not well
described without the presence of a finite frequency chiral
oscillator $\sim 4.5$\,meV. This resonant energy is similar to STM
measurements of the first excited state vortex core energy $\sim
7\,\text{meV}$ above the Fermi energy for optimally doped Bi-2212
 and $\sim 5.5\,\text{meV}$ for
YBCO.\cite{BSCCOVortexEnergyStates7meV} The resonant feature is
also similar to earlier magneto-optical measurements on YBCO of a
hole-like chiral oscillator at $\sim
3\,\text{meV}$.\cite{SteveLihnthesis,ShanLuThesis}

\section{Conclusion}
Measurements of the IR Hall angle on optimally doped Bi-2212 were
performed in fields up to 8T as a continuous function of
temperature at four discrete frequencies ranging from 3 to 22 meV.
Above T$_c$, the IR Hall response characterized by $\omega_H$ is
found to be significantly larger than the value calculated within
a Boltzmann formalism using ARPES measured parameters. This is the
THz manifestation of the well-known anomalous dc Hall effect where
dc-$R_H$ enhancements are much larger than the value expected from
Luttinger's thereom as well as the ARPES+RTA value. These
enhancements as well as the frequency and temperature dependence
of the dc and IR Hall response is well described by a Fermi liquid
theory which incorporates the current vertex corrections produced
by electron-electron interactions mediated by antiferromagnetic
fluctuations.

There exists precursive superconductivity signatures in the
measured IR Hall response and transmission well above T$_c$. The
low frequency data in the superconducting state requires a finite
frequency chiral oscillator $\sim 4.5$\,meV in the conductivity, a
resonance presumably associated with the vortex core states.

\section{Acknowledgement}
This work was supported by the CNAM, NSF (DMR-0030112), and DOE
(DE-AC02-98CH10886). The authors extend their thanks to Andrei B.
Sushkov, Geoff Evans, and Jeffrey R. Simpson for their assistance
in performing the various reported measurements, Matthew Grayson
for supplying the GaAs 2-DEG sample, and Andrea Damascelli and
Giorgio Levy for providing ARPES Tl-2201 data.
%\newpage %Just because of unusual number of tables stacked at end
\bibliography{BSCCOv5}% Produces the bibliography via BibTeX.

%giorgio.levy@physics.ubc.ca
%damascelli@physics.ubc.ca
%m-grayson@northwestern.edu
%ggu@bnl.gov
%jrsimpson@towson.edu
%kon@slab.phys.nagoya-u.ac.jp
%sushkov@umd.edu
%

\end{document}